\documentclass{IEEEcsmag}

\usepackage[colorlinks,urlcolor=blue,linkcolor=blue,citecolor=blue]{hyperref}

\usepackage{upmath}
\usepackage{cite}
\usepackage{soul}
\usepackage{amsmath,amssymb,amsfonts}
\usepackage{algorithmic}
\usepackage{graphicx}
\usepackage{textcomp}
\usepackage{xcolor}
\usepackage{dblfloatfix}
\def\BibTeX{{\rm B\kern-.05em{\sc i\kern-.025em b}\kern-.08em
    T\kern-.1667em\lower.7ex\hbox{E}\kern-.125emX}}
    
\usepackage[separate-uncertainty=true, multi-part-units=single]{siunitx}
\DeclareSIUnit{\bit}{b}
\DeclareSIUnit{\byte}{B}

\usepackage{url}
\makeatletter
\g@addto@macro{\UrlBreaks}{\UrlOrds}
\makeatother

\usepackage{comment}
\usepackage{tikz}

\usepackage{balance}

\jvol{}
\jnum{}
\paper{}
\jmonth{}
\jname{Computer}
\pubyear{2021}

\begin{document}


\fbox{\parbox{\dimexpr\textwidth-\fboxsep-\fboxrule\relax}{
\copyright 2021 IEEE. Personal use of this material is permitted.  Permission from IEEE must be obtained for all other uses, in any current or future media, including reprinting/republishing this material for advertising or promotional purposes, creating new collective works, for resale or redistribution to servers or lists, or reuse of any copyrighted component of this work in other works.
}
}
\thispagestyle{empty} 

\title{Stateful Function-as-a-Service at the Edge}

\author{Carlo Puliafito}
\affil{University of Pisa}

\author{Claudio Cicconetti}
\affil{IIT-CNR}

\author{Marco Conti}
\affil{IIT-CNR}

\author{Enzo Mingozzi}
\affil{University of Pisa}

\author{Andrea Passarella}
\affil{IIT-CNR}


\begin{abstract}
\textcolor{black}{In FaaS, an application is decomposed into functions. When functions are stateful, they typically need to remotely access the state, via an external service. On the one hand, this approach makes function instances equivalent to one another, which provides great resource efficiency. On the other hand, accessing a remote state causes increased delays and network traffic, which makes FaaS less attractive to edge computing systems. We propose to generalize FaaS by allowing functions to alternate between remote-state and local-state phases, depending on internal and external conditions, and dedicating a container with persistent memory to functions when in a local-state phase. We present initial results showing that this simple yet powerful pattern allows to better utilize the available resources, which are scarce on edge nodes, while significantly reducing tail latencies, which is key to enable new applications based on real-time Machine Learning (ML), e.g., in smart vehicles and smart factory scenarios.}
\end{abstract}


\maketitle

\textcolor{black}{\chapterinitial{Monolithic} application design has shown its downsides in terms of scalability, maintainability, and agility. The current trend is to decompose complex applications into small pieces of code called microservices, each focusing on a specific aspect of the overall application. Microservices are typically instantiated within lightweight environments, e.g., containers. Function as a Service (FaaS) leverages microservices (which in FaaS are called \textit{functions}) as a starting point to build enhanced cloud computing systems \cite{VanEykFaaS}. FaaS indeed abstracts the operational logic away from function developers, such that they do not need to care about function deployment, scaling, and lifecycle management. Besides, functions run according to an event-based pattern, and users only pay for what they actually use, with fine granularity.}

\begin{figure*}[t!]
\centerline{\includegraphics[width=36pc]{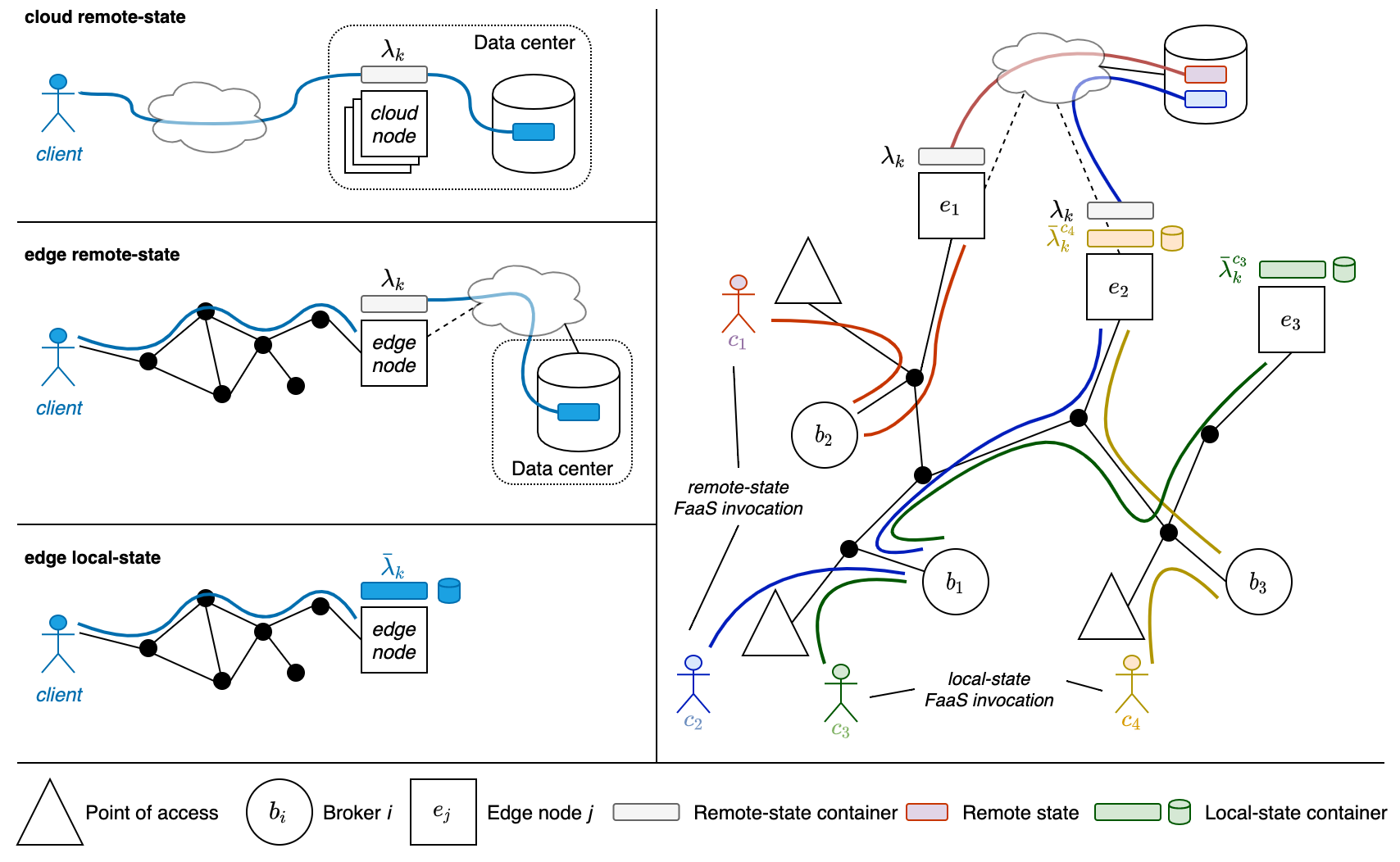}}
\caption{Remote-state vs.\ local-state FaaS invocations at the edge.}
\label{fig:system-model}
\end{figure*}

In this context, consecutive invocations of a function from the same client can be independent from one another or, more often, can form a session with an associated state that must persist across multiple invocations until the session ends \cite{Eismann2021a}. With \textcolor{black}{traditional FaaS for cloud computing systems, functions typically need to remotely access this state at each invocation, via an external service such as a database: we refer to these functions as \textbf{remote-state functions}. This is depicted in the top-left image of Figure~\ref{fig:system-model}, whose notation will be explained in the next section. Following this approach,} different instances of the same remote-state function are equivalent to one another, as they do not retain any state locally \textcolor{black}{(state is download at each invocation, updated, and uploaded again to the external service)}. Therefore, FaaS providers can optimize their infrastructure, transparently to the users, as (i) different users can share the same function instance, (ii) consecutive invocations from the same user can be forwarded to different function instances, and (iii) resources allocated to inactive instances can be freed after a short period of idle time.
The first company to propose a FaaS platform was Amazon with AWS Lambda. Since then, all the top cloud vendors announced their FaaS solutions, e.g., Microsoft Azure Functions, Google Cloud Run, IBM Cloud Functions, and Cloudflare Workers. Open-source platforms, to be executed on private compute infrastructures, are also available, such as Apache OpenWhisk, OpenFaaS, Kubeless, and Knative. \textcolor{black}{Further information on the most prominent FaaS platforms can be found in \cite{FaaStenBrogi}.}

Although it was initially designed for cloud environments, FaaS is gradually drawing interest as a viable option for \textbf{edge computing}, \textcolor{black}{as well \cite{ServerlessEdgeIEEEWirelessCom}}. Edge computing extends the cloud toward the edge of the network, hosting cloud-like services in close proximity to the end users, e.g., on cellular base stations \cite{edgesurvey}. This proximity leads to many advantages, the most important of which is the reduced latency, which is essential to a vast number of emerging applications, such as real-time Internet of Things (IoT), mobile Virtual Reality (VR)/Augmented Reality (AR), and connected vehicle applications~\cite{soft-van,Aslanpour2021}. Big IT companies have started investing in FaaS for edge computing, extending their FaaS platforms toward the edge of the network, for example Amazon IoT Greengrass, Microsoft Azure IoT Edge, and IBM Edge Functions.

Notwithstanding these recent efforts toward FaaS for edge computing, there is still hesitation to widely adopt this novel paradigm. This is due to the cloud-oriented design of FaaS, which does not always suit the distinguishing characteristics of edge applications. The most important design assumption of FaaS that is violated by its expansion toward the edge is that functions access a remote state. \textcolor{black}{In cloud-only environments, this approach affects performance only slightly because both function instances and session state are hosted by servers that are physically located in the same data center. However, when function instances run at the edge (as shown in the center-left image of Figure~\ref{fig:system-model}), accessing a remote state may cause significant service latency and network traffic, at risk of nullifying edge computing advantages.}

To overcome the above limitation, \textbf{local-state functions} are coming into the picture \cite{Lopez2021}. \textcolor{black}{As depicted in the bottom-left image of Figure \ref{fig:system-model}, these functions keep the state locally.}
\textcolor{black}{On the one hand, local-state functions avoid the delays and traffic caused by accessing state from an external storage service. However, on the other hand, they do not experience the same cost-efficiency and flexibility of remote-state functions. Local-state instances are indeed not equivalent to one another, as each is dedicated to a specific user or application session, for which it provides data access in a private and persistent manner. Besides, local-state function instances are not triggered on demand; instead, they are long-running to retain state across invocations.} 
The following are examples of local-state functions in commercial FaaS platforms: (i) Microsoft entity functions \cite{microsoftdurable}; (ii) Cloudflare durable objects \cite{cloudflaredurable}; and (iii) Amazon long-lived functions \cite{amazonlonglived}.

Today, the choice on whether a given function should follow a remote-state vs.\ local-state pattern is made at design time, and migrating from one pattern to another in production can be very expensive, since it involves changing the set of employed services (adapting to new Application Programming Interfaces (APIs), switching contracts, using a different Software Development Kit (SDK)).
What is worse, during development the programmer may not even know whether the logic of the code they are implementing will be executed at the edge or in the cloud, so making an informed choice could be just impossible.

In this work, we advocate that such a dichotomy between remote-state and local-state should not exist, but rather a function in a FaaS environment should be able to \textbf{adapt dynamically}, i.e., changing its behavior from remote-state to local-state and \textit{vice versa}, depending on both internal and external factors. \textcolor{black}{This approach would relieve the developer from the risk of making an uninformed decision. Besides, it would let the FaaS provider carry out run-time optimizations, e.g., to increase resource efficiency. Finally, it would benefit applications with requirements that dynamically change over time.}   

We first present our proposal at a high level. To showcase the potential advantages of dynamically changing the nature of functions in a FaaS environment, we then report initial results exploring the main trade-offs involved in this approach.
Next, we describe two practical use cases of business interest that can benefit from our idea. We then report the essential related work in the field. Finally, we conclude the paper and outline the further research directions originating from our proposition.

\section{STATEFUL FAAS AT THE EDGE}\label{sec:model}

We illustrate our proposal within a system model that abstracts the specific and technical details of a real edge system, which consists of the following elements:

\begin{itemize}
\item \textit{clients}, wishing to invoke functions $\lambda_i$ of a given type (or application) $i$: consecutive invocations of a function from the same client are called a \textit{session}, which has an associated \textit{state} that is expected to persist until the session ends;

\item \textit{brokers}, representing entry points of the system for the clients, i.e., the latter invoke their functions on the broker, which then delegates the actual execution of the function to a \textit{worker} (i.e., a container) in the edge network of matching type;

\item \textit{workers}, handling function invocations and hosted by containers. Remote-state containers are instances of remote-state functions, and therefore rely on an external service, possibly located in the cloud, to access the session state. On the other hand, local-state functions get instantiated in local-state containers, which are associated to a specific session and keep any state required locally.
\end{itemize}

Edge nodes may host any combination of workers and brokers. In this work, we indicate the remote-state function of type $k$ as $\lambda_k$, whereas local-state function of type $k$ is $\bar\lambda_k$. The considered system works in mixed remote-state + local-state conditions. This can be true both from the point of view of different functions of type $h \ne k$ and for the same function $k$. \textcolor{black}{The right image of Figure \ref{fig:system-model} depicts an example of such a mixed behavior. Function $\lambda_k$ is invoked by four different clients $c_1,\ldots,c_4$. For clients $c_1$ and $c_2$, $\lambda_k$ is instantiated in remote-state containers.}  
%
These instances of $\lambda_k$ are indistinguishable from one another, and in fact can be scaled up and down (also to zero instances) by the underlying container orchestration mechanism.
The brokers need only to know the locations of all (or a subset) of the containers and can then implement all sorts of decentralized load balancing as discussed in \cite{9193994}.
%
For instance, the invocation from $c_1$ is forwarded to the $\lambda_k$ instance hosted on $e_1$. However, the next invocation could be equally forwarded to the instance on $e_2$. This gives the system flexibility in resource scheduling. Yet, this solution has two main disadvantages: (i) the response time also includes the time required for the function instance to synchronize the state on the external service; (ii) network traffic is generated as a consequence of state synchronization.
%

\textcolor{black}{On the other hand, clients $c_3$ and $c_4$ use local-state containers $\bar\lambda_k^{c_3}$ and $\bar\lambda_k^{c_4}$, respectively.} Local-state containers are more bandwidth-efficient and do not incur in the same latency associated to remote-state containers, as explained above.
However, they do not enjoy the same orchestration flexibility, either. Rather than maintaining a pool of shared containers sufficient to serve the current number of active clients, one local-state instance must exist in the edge network for each session. For illustration purposes, in the example we assume without loss of generality that every client has exactly one session. 
%
%
Therefore, when a broker receives a function invocation, it must forward it to the container specific to that client.
%
Also, if the platform wants to move a local-state container to another edge node, a live migration is required to transfer the state, as well as the image: this has a cost in terms of network traffic and creates a period while the container is unavailable (i.e., downtime).

\begin{figure*}[t!]
\centerline{\includegraphics[width=36pc]{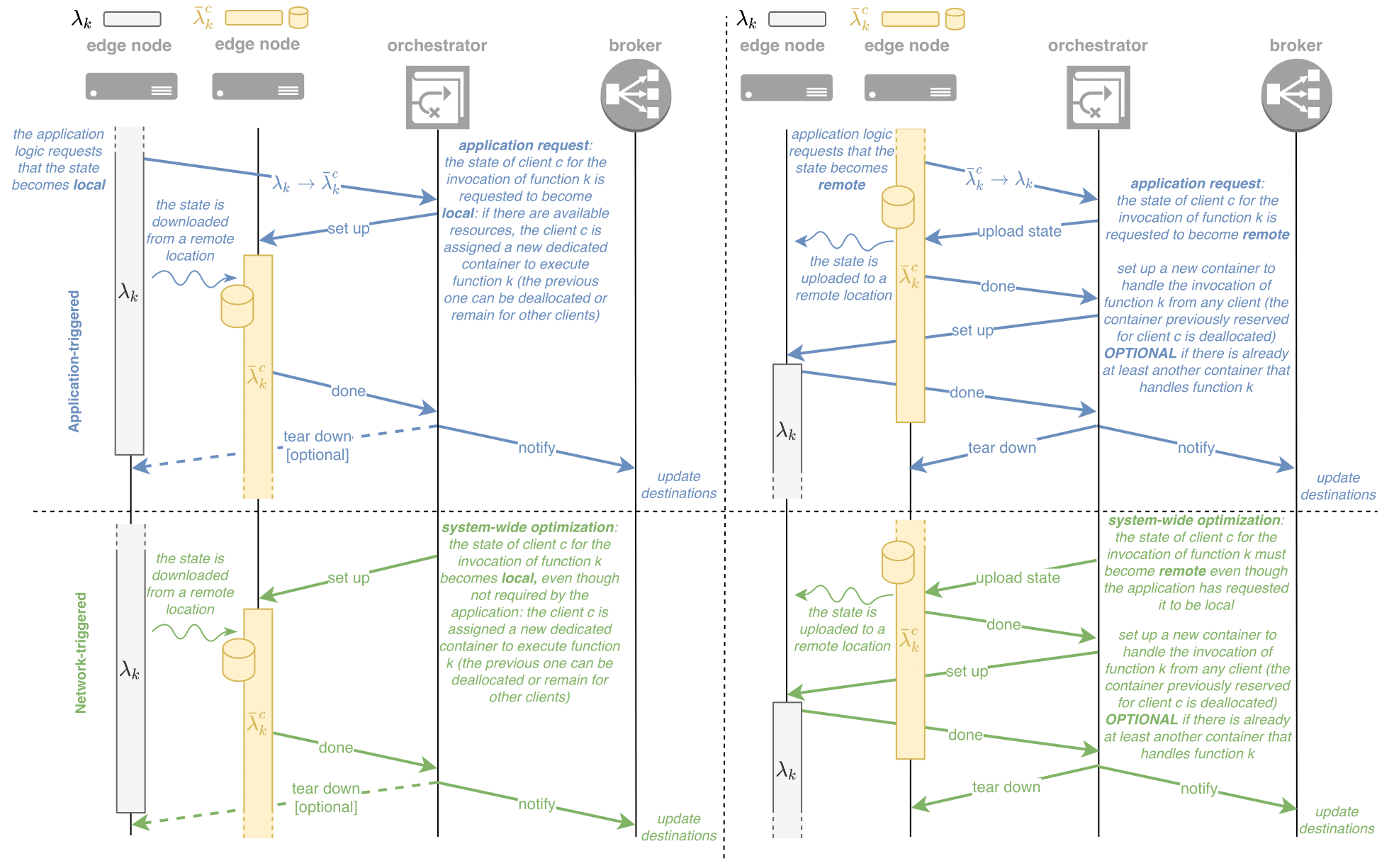}}
\caption{Sequence diagram of transition of a worker from remote-state to local-state (left) or local-state to remote-state (right), as triggered by the application (top) or by the network (bottom).}
\label{fig:sequence}
\end{figure*}


\textcolor{black}{The example above shows the limitations of a system where functions are statically instantiated as either remote-state or local-state. Any of the two patterns presents some drawbacks, indeed.} The main contribution of this work is proposing a paradigm where functions are able to adapt dynamically to unpredictably changing conditions, by changing behavior from remote-state to local-state and \textit{vice versa}.

To support this paradigm, the most natural way would be that the developer of a function $\lambda_k$ provides two versions (i.e., container images) with the same application logic: a remote-state version $\lambda_k$ and a local-state version $\bar\lambda_k$. Besides, the developer of the function is expected to implement some means to download the state locally from the external service in use and to upload a local state to the external service intended to be used \textcolor{black}{(which is true also in traditional FaaS systems).}
%
%
%
The details on the application internals, such as the programming language it uses or which external services are used (and how), do not need to be disclosed to the FaaS platform.

\textcolor{black}{We believe that this dynamic transition from remote-state to local-state, and \textit{vice versa}, may be useful (and therefore be triggered) for two main purposes. One is to allow the service provider to perform run-time optimizations, e.g., increase resource efficiency. We refer to this type of transition as \textbf{network-triggered transition}, since it is activated by the platform. Alternatively, another purpose is to accommodate applications having requirements that dynamically change over time. In this case, we talk about \textbf{application-triggered transition}, as it is the application to request it.}


\textcolor{black}{Figure~\ref{fig:sequence} presents the possible sequence diagrams of the transitions of a worker. Specifically, transitions to local-state are shown on the left, whereas transitions to remote-state are depicted on the right. In a similar way, application-triggered transitions are at the top in figure, while network-triggered transitions are at the bottom. As shown, application-triggered and network-triggered transitions work in the same way, a part from the initial triggering event, which is different in the two cases.}

\textcolor{black}{Let us start with a transition from remote-state to local-state behavior. Initially, client $c$ uses remote-state instances of function $\lambda_k$. Then, after checking available resources, the system orchestrator sets up a local-state container $\bar\lambda_k^{c}$ and assigns it to client $c$. When $\bar\lambda_k^{c}$ starts, it first downloads the session state of client $c$ from the external service (where it has been previously uploaded by the remote-state instance, as per its normal working) and stores it locally. It then notifies the system orchestrator, which therefore informs the broker to update the record for client $c$. As a result, any future function invocation of client $c$ is forwarded to $\bar\lambda_k^{c}$ by the broker. The remote-state instance of function $\lambda_k$ that was used by $c$ in its last invocation can be either deleted or remains active for other clients.}

\textcolor{black}{For what concerns transition to remote-state (see Figure \ref{fig:sequence} on the right), the starting point is that any invocation from client $c$ is forwarded by the broker to the dedicated instance $\bar\lambda_k^{c}$. When the triggering event for the transition is fired, the system orchestrator requires $\bar\lambda_k^{c}$ to upload the session state to the external service. When this is done, the system orchestrator might decide to create a new instance of remote-state function $\lambda_k$ or use the ones that already exist, if any. The system orchestrator then tears down $\bar\lambda_k^{c}$ and notifies the broker to update the record for client $c$. Any future invocation from $c$ can be forwarded by the broker to any remote-state instance of function $\lambda_k$.}

In the next section we show, with the help of a simple analytical model, that the benefits of breaking the dichotomy remote-state/local-state can be significant.

\section{EVALUATION}\label{sec:eval}

In this section, we report the results obtained with a simple analytical model, with the purpose of showing the significant advantages that can be expected by applying the proposed approach and highlighting key open research directions accordingly.

We consider two scenarios.
In the first scenario, a number of independent clients, with same characteristics, issue function invocations towards a pool of identical containers at the edge.
To keep the model simple, both the inter-time between consecutive invocations and the function execution time are distributed exponentially: when a function is treated as local-state, then its dedicated container takes on average \SI{1}{\second} to execute the function; on the other hand, remote-state functions require on average \SI{3}{\second} to be dispatched by the container, because of the overhead to copy back and forth the application state as discussed previously.

\begin{figure}[t!]
\centerline{\includegraphics[width=18.5pc]{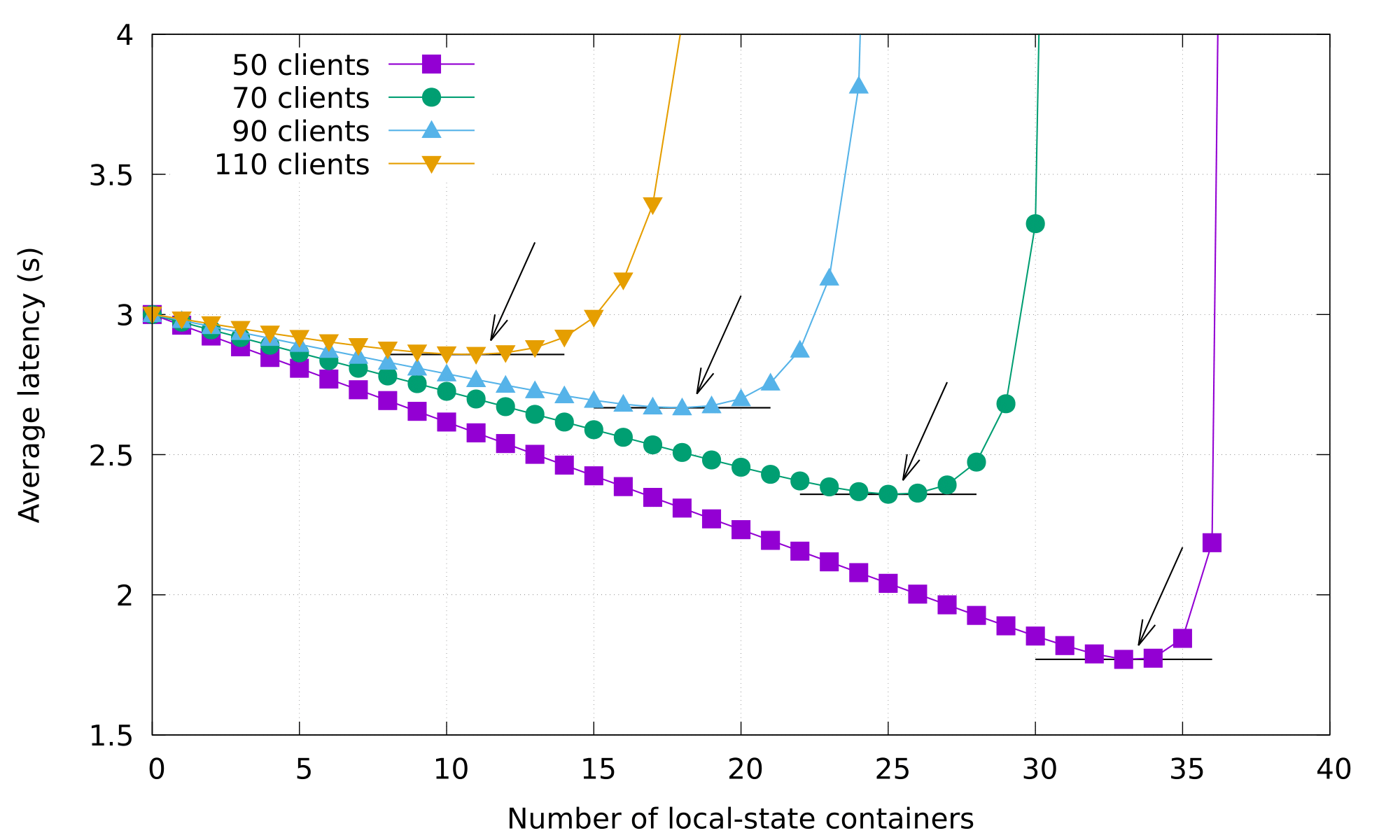}}
\caption{Average latency vs.\ number of containers assigned to local-state functions, for 50, 70, 90, and 110 clients. The arrows point to the minimum of each curve.}\label{fig:latency}
\end{figure}

We assume that the number of containers provisioned is fixed and equal to 40 and that clients perform 4.5 function calls per minute.
Even in this simple scenario, the service provider has one degree of freedom that it can use to optimize the system performance: by employing the network-triggered transition pattern (as in the bottom of Figure \ref{fig:sequence}) it can force some of the functions to be treated as either remote-state or local-state.
A question the service provider might ask is: \textit{how many containers should be dedicated to local-state functions at any time, provided that there are not enough for all the active ones?}

Intuitively, there is the following trade-off: the higher the number of local-state functions, which enjoy a smaller delay due to i) lack of competition at container level and ii) the local availability of the state, the lower the containers available for shared used by the remote-state clients, which will suffer from increasing scarcity of resources.
The trade-off is shown in a quantitative manner in Figure \ref{fig:latency}, which plots the average latency (considering both the local-state and the remote-state functions, weighted on their respective cardinalities) as the number of local-state containers increases:
after an initial period where dedicating containers to local-state functions is beneficial, a minimum is reached after which the average delay increases again sharply until the system becomes quickly unstable, i.e., the service queues grow indefinitely.
Such a behavior happens irrespective of the number of clients, but is more pronounced with a higher population.
The results strongly suggest two key properties. First, a dynamic management allowing to switch between local- and remote-state functions can lead to very significant performance advantages over static configurations, and configuring the system at the optimal operating point is fundamental. Second, the optimal operating point varies significantly as a function of the involved parameters (number of clients, in this specific example), and thus trivial optimization approaches may not be sufficient. Both properties indicate that the role of an orchestrator taking non-trivial run-time decisions is crucial to achieving optimal performance. 


\begin{figure}[b!]
\centerline{\includegraphics[width=18.5pc]{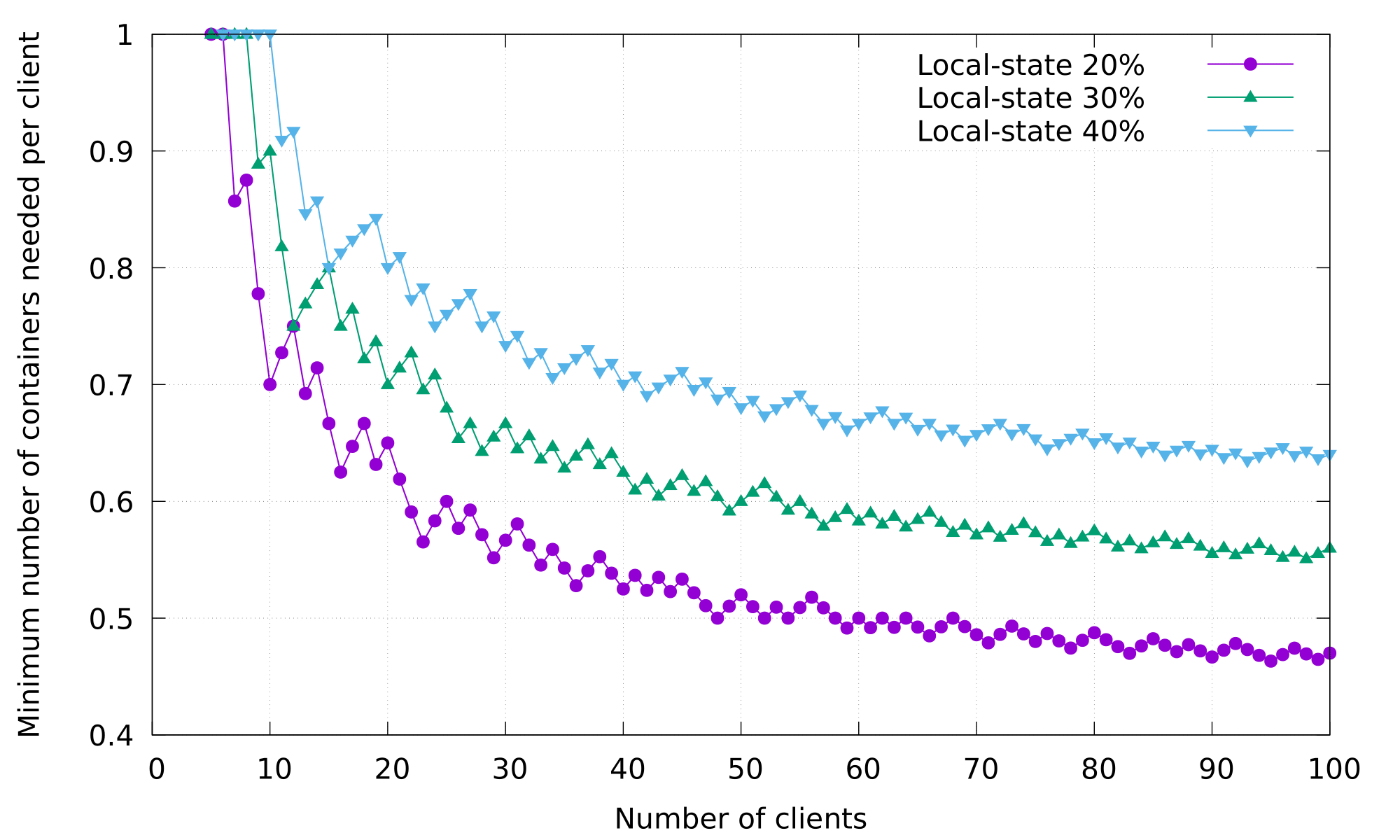}}
\caption{Minimum number of containers per client required to guarantee that no more than 1\% of the functions requesting to be served as local-state are served instead as remote-state, with increasing number of clients. We also varied the percentage of time a function requests to be served as local-state (20\%, 30\%, and 40\%).}\label{fig:provisioning-multi}
\end{figure}

In the second scenario, we consider the case of application-triggered transitions (as in the top of Figure \ref{fig:sequence}): the client applications decide by themselves whether they would prefer their functions to be served local-state or they can accept being treated as remote-state with no penalty for the user.
We model the transitions between the need of being served in a local- vs.\ remote-state manner as a two-state Markov chain, with different combinations of the transition probabilities such that the percentage of time a client application requests its function to be served as local-state is 20\%, 30\%, and 40\%.
We then asked ourselves the following question, again from the point of view of the service provider: \textit{given a number of clients, how many containers should be provisioned to make sure that the system is stable (i.e., buffers do not grow indefinitely) and the probability that a given client requesting its function to be local-state is treated as remote-state instead (due to a shortage of containers) is small enough (e.g., less than 1\%)?}

The answer is plotted in Figure \ref{fig:provisioning-multi} for a variable number of clients.
The plot shows the minimum number of containers that are required to match the service provider conditions per client.
For instance, with 10 clients and for client applications requesting local-state 20\% of the time, the plot tells us that we need at least 0.7 containers/client, i.e., 7 containers.
These results can thus be used to provision the number of containers, in accordance with the Service Level Agreements (SLAs) and other system constraints.
It is interesting to note that, as the number of clients increases, all the curve stabilize around constant values (20\%: 0.47; 30\%: 0.55; 40\%: 0.64), which depend on the transition rates of the applications between local and remote states, as well as the other load characteristics.
Therefore, such an analysis, extended to take into account more realistic conditions and the real characteristics of the target deployment, could provide simple but precious rules for the provisioning of a stateful FaaS system at the edge (in this scenario, for example with 20\% local-state, the rule would be: make sure that the number of containers is at least half the number of clients).

\section{USE CASES}\label{sec:usecases}

Our vision of FaaS for edge computing can empower emerging use cases in a resource-efficient and performing way. The applications that most benefit from our solution are stateful ones having requirements that dynamically change over time. When the application has strict requirements (e.g., latency), maintaining the state locally should be preferred. However, this requires the container to be a dedicated and long-running resource, resulting in a non-negligible cost. As a result, when application requirements are looser, it may be more convenient to access a remote (e.g., cloud-hosted) state and let more users share the same container. This second approach is more resource- and cost-efficient but degrades performance due to high latency, queuing, and increased network traffic. In what follows, we briefly describe some use cases that present the above dynamic characteristics and may therefore take advantage of our idea.

\subsection{Smart vehicles}

\begin{figure*}
\centerline{\includegraphics[width=36pc]{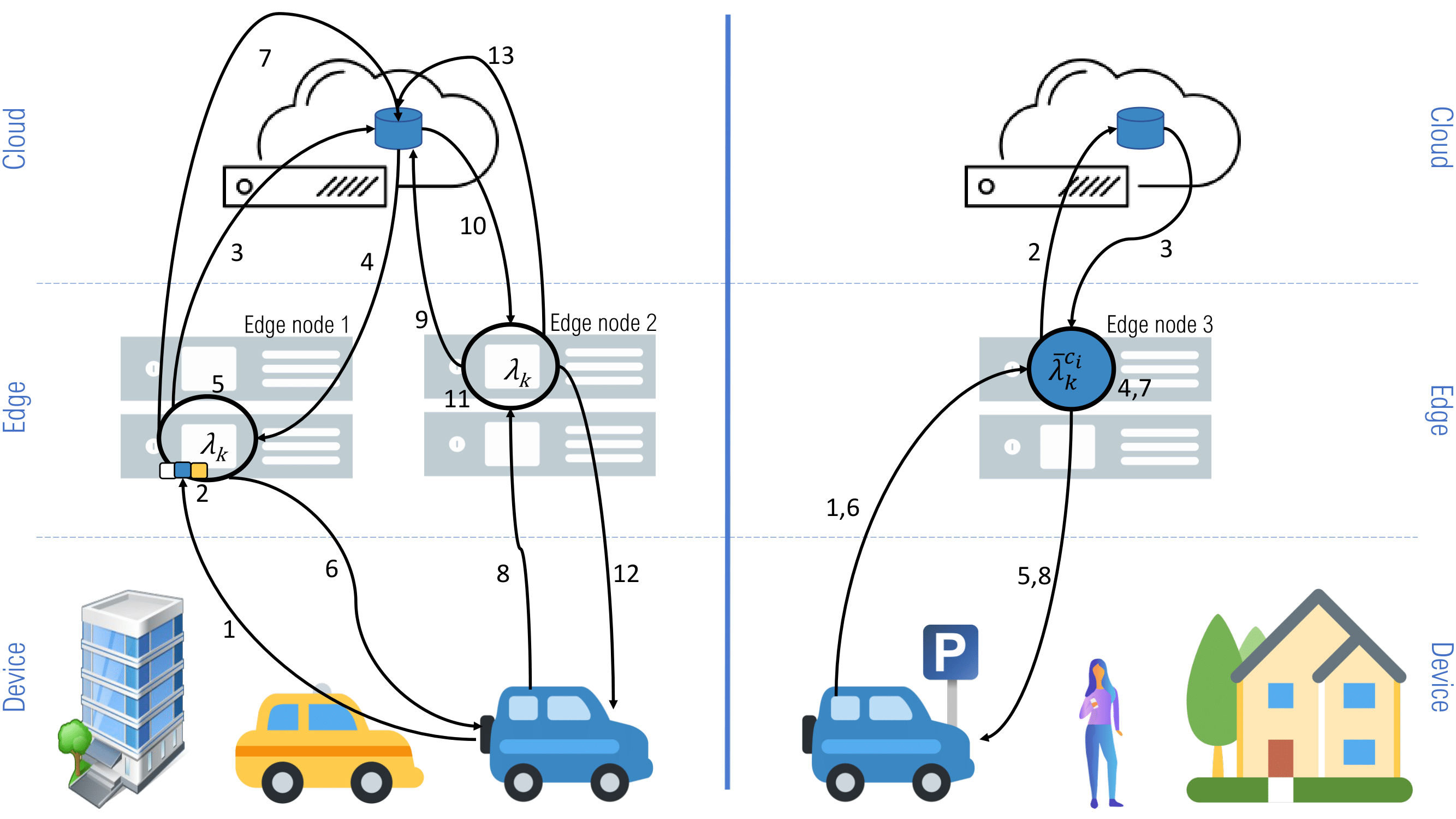}}
\caption{On the left, remote-state containers run a regular driving logic; on the right, a local-state container runs an autonomous parking logic.}
\label{fig:usecase}
\end{figure*}

Autonomous vehicles and Advanced Driver-Assistance Systems (ADAS) are gaining momentum as a way to enhance safety and reduce traffic congestion.
Our use case takes inspiration from \cite{vehicleparking} and is depicted in Figure \ref{fig:usecase}. Let us suppose that a driver takes her blue car to travel back home from the office, and that the car uses FaaS with a function that is in charge of assisting the driver. During regular driving, ADAS limits to speed control and steering of the vehicle, which have loose latency and throughput requirements --- \SI{1000}{\milli\second} and \SI{0.2}{\mega\bit\per\second}, respectively \cite{huawei5g} --- and the function runs as remote-state $\lambda_k$. As shown in the left side of Figure \ref{fig:usecase}, the first invocation of the function is forwarded to a container running on edge node 1 (step 1). The invocation is queued because the container has been previously invoked by the yellow car (step 2). When the blue car can be served, the $\lambda_k$ container retrieves the session state from a remote storage (steps 3 and 4), computes the response (step 5), forwards it to the user's car (step 6), and updates the remote storage with the new session state (step 7). In step 8, $\lambda_k$ is again invoked. However, this time, a container running on edge node 2 is invoked. As illustrated, the $\lambda_k$ container has to access the remote storage again to read and write the session state.
When the user reaches home, as illustrated in the right side of Figure \ref{fig:usecase}, the function enters the autonomous parking routines, which have tighter requirements ---  \SI{10}{\milli\second} and \SI{100}{\mega\bit\per\second}, respectively \cite{huawei5g}. As a result, the function changes from remote-state to local-state: as the logic is invoked in step 1, the session state is retrieved from the remote storage to instantiate the function as a local-state container $\bar\lambda_k^{c_i}$, and all the next invocations of the function are forwarded to the same instance and do not need any access to the remote storage (e.g., steps 6-8).

\subsection{Smart factory}

This use case is based on \cite{bosch} and \cite{robotcontrol}. In a smart factory, a robotic arm periodically sends information on its operation (e.g., positioning, temperature of the CPU, temperature of the case) to a monitoring service. Under normal conditions, the monitoring service can be a remote-state function invoked on demand. This allows more robotic arms in the manufacturing plant to share the same container, thus saving resources. However, when the monitoring service predicts an abnormal functioning, the sampling frequency at the robotic arm increases, and the monitoring service is deployed as a local-state container dedicated to the malfunctioning arm. This is indeed necessary for prompt reactions, e.g., emergency stops, that avoid damages to the arm and people in the vicinity. \textcolor{black}{We nonetheless highlight that the remaining (faultless) robotic arms can continue to invoke the monitoring service using (shared) remote-state containers. This use case, as well as the previous one, shows how our approach could meet dynamically changing requirements of applications, while providing resource efficiency. For instance, recalling insights from the previous section, the service provider can provision a total number of containers that is lower than the number of robotic arms and still meet the dynamic requirements of each of them.}

\section{RELATED WORK}\label{sec:soa}

Some scientific papers have addressed already the problem of state handling in FaaS. In \cite{baresi}, authors present a prototype implementation of a FaaS platform for edge computing based on Apache OpenWhisk. They propose local-state functions, such that each instance keeps its state locally and is associated with a unique session token that distinguishes it from other instances. 
In \cite{skippy}, authors propose Skippy, a container scheduling system that optimizes the placement of remote-state functions in the cloud-edge continuum based on (potentially) conflicting aspects such as: (i) location of edge storage nodes that are accessed by functions; (ii) location of container base image registries; (iii) hardware capabilities of edge nodes; (iv) location of node (either edge or cloud). 
Cloudburst \cite{cloudburst} is a solution designed for cloud data centers, with the main goal of improving performance of storage access by remote-state functions. Authors assume a composition of functions that share a state and make up a complex application. A centralized key-value store is shared among the functions. However, accessing this store involves high latency. Therefore, Cloudburst introduces a data cache on each compute node, which is accessible by all and only the function instances running on that node. 
These works show that there is a growing interest on the topic of stateful functions in FaaS, but none of them have considered that a single function could dynamically adapt its nature, remote-state or local-state, depending on the environment, which is our key value proposition.

\section{CONCLUSIONS}\label{sec:conclusions}

In this paper, we have considered the execution of stateful functions at the edge, which is an emerging necessity especially for real-time IoT applications.
We have defined a generic model where functions can execute either in a stateful container dedicated to the given application instance (\textit{local-state functions}), or in a pool of stateless containers with the need to access the state of the application instance in a remote facility (\textit{remote-state functions}).
We have illustrated two example use cases, i.e., smart vehicles and smart factory, to show that the system under study has potential impact on applications of high economic and social impact.
Using a simple model we have then shown that there are interesting performance trade-offs, in terms of (e.g.) the application latency and the amount of resources used.

Our contribution is merely intended to raise awareness on the potential of unleashing dependence from the state at design/development stage. Such an approach shows significant potential in terms of performance improvements over static designs and opens several research challenges on how to optimize the system operation (\textit{maximizing clients' revenue under constrained resources? minimizing the system costs under minimum target application performance?}) by executing the remote-state $\leftrightarrow$ local-state transition depending on the internal status of the application (\textit{training vs.\ inference for a continual learning ML application; regular operation vs.\ alarm condition for a monitoring system; etc.}) and the edge run-time environment (\textit{load of edge nodes and their amount of memory/storage available; instantaneous network traffic in the edge; amount of outbound traffic; function response times; etc.}).

\section*{Acknowledgements}

This work was partially supported by the European Commission (Horizon 2020) in the framework of the project "Multimodal Extreme Scale Data Analytics for Smart Cities Environments (MARVEL)" under Grant Agreement no. 957337, and by the Italian Ministry of Education and Research (MIUR) in the framework of the CrossLab project (Departments of Excellence).

\balance



%

\begin{IEEEbiographynophoto}
{Carlo Puliafito}%
has a PhD in Smart Computing jointly from the Universities of Florence and Pisa (Italy). He is a research fellow at the University of Pisa. His research interests include edge computing and Internet of Things. Contact him at carlo.puliafito@ing.unipi.it.
\end{IEEEbiographynophoto}

\begin{IEEEbiographynophoto}{Claudio Cicconetti}%
has a PhD in Information Engineering from the University of Pisa
(Italy) and he is a Researcher at IIT-CNR.
He is interested in serverless edge computing
and Quantum Internet architecture and protocols. Contact him at c.cicconetti@iit.cnr.it.
\end{IEEEbiographynophoto}

\begin{IEEEbiographynophoto}{Marco Conti}%
is the Director of IIT-CNR.
He is interested in design, modelling, and performance evaluation
of computer and communications systems, and their use for decentralized
solutions for self-organizing networks. Contact him at m.conti@iit.cnr.it.
\end{IEEEbiographynophoto}

\begin{IEEEbiographynophoto}{Enzo Mingozzi}%
is a Full Professor with the Department of Information Engineering, University of Pisa, Italy.
He received a PhD in Computer Systems Engineering from the same university. His research interests span several areas, including resource optimization in wireless and wired networks, mobile Edge/Fog Computing and the Internet of Things. Contact him at: enzo.mingozzi@unipi.it.
\end{IEEEbiographynophoto}

\begin{IEEEbiographynophoto}{Andrea Passarella} has a PhD in information engineering awarded by the University of Pisa
(Italy). He is a Research Director at IIT-CNR and Head of the Ubiquitous
Internet Group.
He is interested in content-centric networks, Internet of People,
and explainable AI. Contact him at a.passarella@iit.cnr.it.
\end{IEEEbiographynophoto}




\end{document}